\def\civ{{\sc{Civ}}$\lambda$1549\/} 
 
\def\ltsima{$\; \buildrel < \over \sim \;$} 
\def\simlt{\lower.5ex\hbox{\ltsima}} 
\def\hb{{\sc{H}}$\beta$\/} 
\def\hbbc{{\sc{H}}$\beta_{\rm BC}$\/} 
\def\hbvbc{{\sc{H}}$\beta_{\rm VBC}$\/} 
\def\hbnc{{\sc{H}}$\beta_{\rm NC}$\/} 
 
\def\ne{n$_{\rm e}$\/} 
 
\def\rfe{R$_{\rm FeII}$} 
\def\feiiq{\rm Fe{\sc ii }$\lambda$4570\/} 
\def\REF{\par\noindent\hangindent 20pt} 
\def\msol{M$_\odot$\/}

\def\ltsima{$\; \buildrel < \over \sim \;$} 
\def\simlt{\lower.5ex\hbox{\ltsima}}            
\def\gtsima{$\; \buildrel > \over \sim \;$} 
 
\def\simgt{\lower.5ex\hbox{\gtsima}}            

\def\civ{{\sc{Civ}}$\lambda$1549\/}

\def\cmt{cm$^{-2}$\/} 
\def\cm3{cm$^{-3}$\/} 
\def\hb{{\sc{H}}$\beta$\/} 
 
\def\hbbc{{\sc{H}}$\beta_{\rm BC}$\/} 
\def\hbnc{{\sc{H}}$\beta_{\rm NC}$\/}

\def\o4363{{\sc{[Oiii]}}$\lambda$4363\/}

\def\feii{{\sc{Feii}}$_{\rm opt}$\/} 
\def\fe{{\sc{Fe}}\/} 
\def\heii{{\sc{Heii}}$\lambda$4686\/} 
\def\heiivbc{{\sc{Heii}}$\lambda$4686$_{\rm VBC}$\/} 
\def\heiibc{{\sc{Heii}}$\lambda$4686$_{\rm BC}$\/} 
\def\fe76087{{\sc [Fe vii]}$\lambda$6087\/} 
\def\oiii{{\sc [Oiii]}$\lambda\lambda$4959,5007} 
 
\def\kms{km~s$^{-1}$} 
\def\ergss{ergs s$^{-1}$\/} 
 
 \documentclass[preprint]{aastex} 
 
\shorttitle{PG 1416--129} \shortauthors{Sulentic et al.}

\begin{document}

\title{The Demise of the Classical BLR in the Luminous Quasar 
PG1416--129}

\author{J. W. Sulentic\altaffilmark{1}, P. Marziani\altaffilmark{2}, 
T. Zwitter\altaffilmark{3}, D. Dultzin-Hacyan\altaffilmark{4} 
and M. Calvani\altaffilmark{2}} 
 
\altaffiltext{1}{Department of  Physics and Astronomy, University 
of Alabama, Tuscaloosa, AL 35487; giacomo@merlot.astr.ua.edu} 
 
\altaffiltext{2}{Osservatorio Astronomico di Padova, vicolo 
dell'Osservatorio 5, I-35122 Padova, Italy; marziani@pd.astro.it; 
calvani@pd.astro.it} 
 
\altaffiltext{3}{Department of Physics, University of Ljubljana, 
Jadranska 19, 1000 Ljubljana, Slovenia; tomaz.zwitter@uni-lj.si} 
 
\altaffiltext{4}{Instituto de Astronomia, UNAM, Mexico, DF 04510, 
Mexico; deborah@astroscu.unam.mx} 
 
\begin{abstract} 
New observations of the broad-line quasar PG1416--129 reveal a 
large decline in its continuum luminosity over the past ten 
years. In response to the continuum change the ``classical'' broad 
component  of \hb\ has almost completely disappeared (a $\times$10  
decrease in flux). In its place 
there remains a redshifted/redward asymmetric very broad emission 
line  component. The significance of  this change is multifold. 
(1) It confirms the existence  of a distinct redshifted Very 
Broad Line Region (VBLR) component that persists after the demise 
of the broad component and that is frequently observed, along with 
the broad component, in radio-loud sources. (2) The smaller 
($\times$2) intensity change in the \hb\ very broad component 
supports the previously advanced idea that the VBLR is physically 
distinct and likely to arise in an optically thin region close to 
the central source. (3) The presence of a strong very broad 
component  in the radio-quiet quasar PG1416--129 reinforces the 
notion that such ``population B" quasars share similar 
spectroscopic (and hence geometrical and kinematical) properties 
to radio-loud sources. (4) AGN can show broad, very broad, or 
both line components simultaneously, making statistical 
comparisons of source profile widths difficult. (5) The 
interpretation, in reverberation studies, of the presence or lack 
of  correlated response in broad line wings will be affected by 
this composite BLR/VBLR structure. 
\end{abstract}

 
\section{Introduction}

We present new spectroscopic observations of the radio quiet
quasar PG 1416--129 that reveal  a significant 
decrease in optical continuum intensity and the virtual 
disappearance of the ``classical'' broad component of \hb\ 
(hereafter indicated as \hbbc) last observed 10 years ago. There 
remains a strong, very-broad (indicated as \hbvbc) and redshifted 
component that has declined much less dramatically in response to 
the continuum change. In \S 2  we describe the new 
observations, data reduction procedures and a comparison of the old 
and new spectra  for this source. 
This is followed  (\S \ref{impl}) by a discussion of physical and 
phenomenological implications of the spectra change in PG 1416--129.

\section{Observations and Analysis \label{obs}} 
 
New spectra of PG 1416--129 (z=0.12928$\pm$0.00007 from [OIII] 
$\lambda$5007) were obtained on the 2.12-m telescope 
of the Osservatorio Astronomico Nacional at San Pedro Martir, Baja 
California, Mexico on May 9 2000 between 7:28 and 8:32 UT, under 
photometric sky conditions. A Boller \& Chivens spectrograph was 
equipped with a 600 l/mm grating with a Tektronix TK1024AB CCD 
chip as the detector (2 arcsec slit width). Suitable standard 
stars were observed at a similar airmass. The spectrum covers the 
range $\lambda \lambda 4438-6537$\AA\ with an effective spectral 
resolution of 3.5 \AA\ FWHM. It was reduced with standard IRAF 
procedures. Line properties were derived after normalization of 
the two spectra to the \oiii\ flux in a spectrum obtained on 
February 17, 1990 (Boroson \& Green 1992: hereafter 
BG92). 
 
\begin{figure} 
\plotone{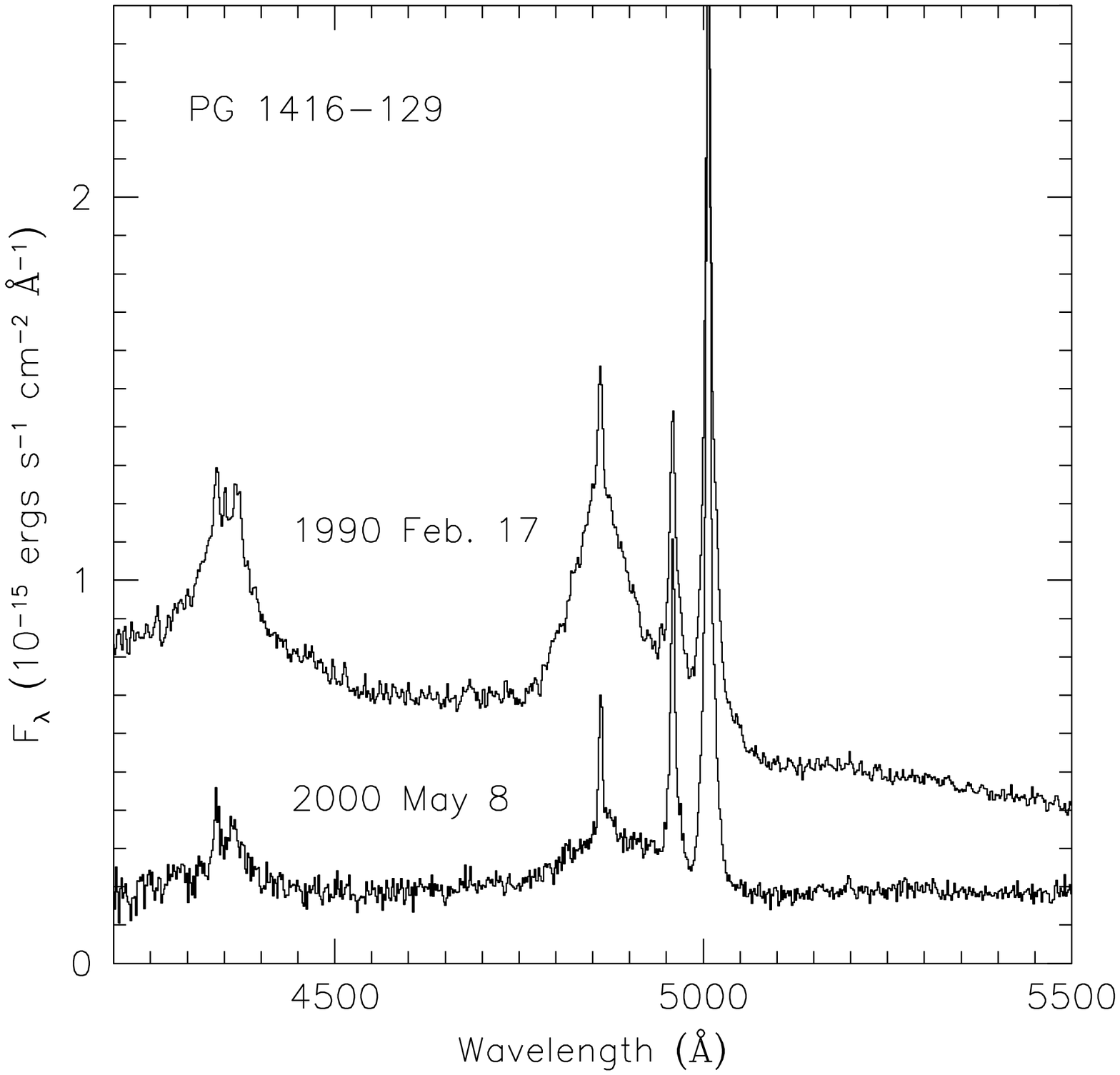} 
\figcaption[fig01.eps]{Spectra of the 
H$\beta$\ spectral region of PG 1416--129 on Feb. 17, 1990 and on 
May 8, 2000. Horizontal scale is rest-frame wavelength in \AA, 
vertical scale is observed specific flux (as in BG92). The two spectra have 
been normalized to the  \oiii\ flux measured in 1990. 
\label{fig1}} 
\end{figure}

\begin{figure} 
\plotone{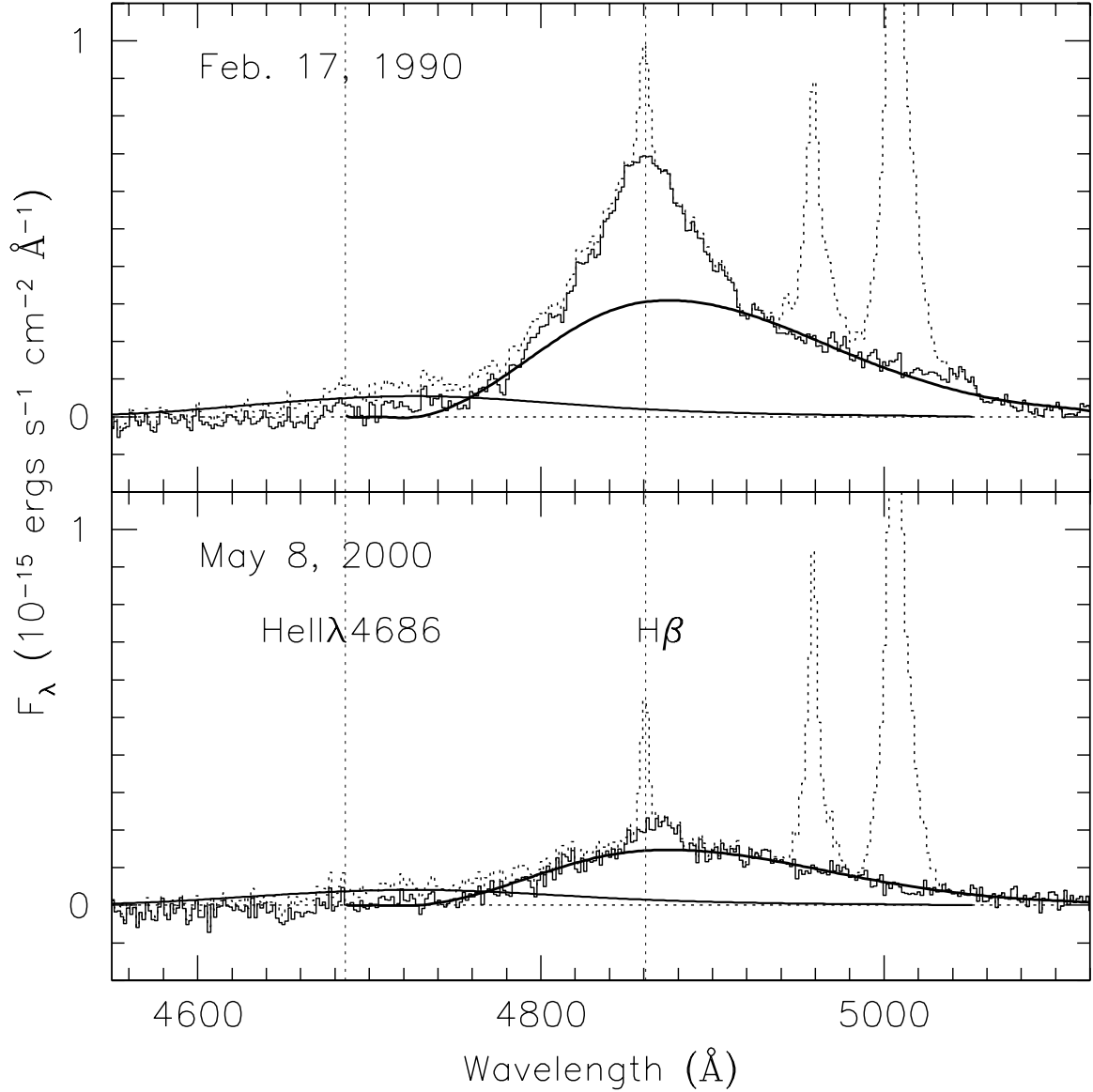} 
\figcaption[fig02new.eps]{Continuum 
subtracted H$\beta$ line profiles on 1990 Feb. 17 (upper panel) 
and on 2000 May 8 (lower panel). The dotted line traces the 
original spectrum after \feii\ subtraction (\feii\ subtraction is 
however modest and has a negligible effect on the \hb\ profile). 
The solid thin line represents the broad line spectrum, after 
narrow line and \heii\ subtraction. The smooth, thick solid line 
in the lower panel is the result of a low order spline fitting to 
the Very Broad Component of \hb; in the upper panel, the fit 
result on the 2000 May 8 spectrum  has been scaled by a factor 
2.1 to match the 1990 Feb 17 observation. The  \hbvbc\ spline fit 
has been scaled and shifted to account also for the \heii\ profile 
(smooth thin lines). \label{fig2}} 
 
\end{figure}

Figure \ref{fig1} shows the 4200-5500$\rm{\AA}$ rest frame region 
in both the 1990 and 2000 spectra. After subtraction of the 
underlying continuum (fit with a low order polynomial; 
contamination from the light of the host galaxy is negligible), 
the \hb\ profiles were cleaned of \oiii, \hbnc, and \heii\ 
emission (for details see Marziani et al. 1996). Observed line 
flux, rest-frame equivalent width and FWHM values derived from the 2000 May 
8 spectrum are given in Table \ref{tab1}. Errors are 20\%\ (2$\sigma$)
for the flux and equivalent width and 10\%\ (2$\sigma$) for the FWHM. 
\feii\ emission is weak in the 1990 spectrum (BG92 estimated W(\feiiq)= 31 
\AA) and weaker in the new spectrum as expected if it correlates with 
the strength of the classical BLR (e.g. BG92). 
We 
subtracted a scaled and broadened I Zw 1 template ($F_\lambda \le 
1 \times 10^{-14}$ \ergss \cmt) from the 2000 May 8 spectrum in 
order to remove \feii\ contamination. 
\heii\ emission may have a more significant effect on the blue 
side of the \hb\ profile if a significant \heii\ VBC exists in 
our new spectrum.  After several empirical attempts, the 
smoothest residuals were obtained by using the \hbvbc\ measured 
from our new spectrum as a template to subtract suspected broad 
\heii\ emission. This subtraction assumes a much stronger \heii\ 
emission than the estimate given in BG92 (W(\heii)/W(\hb) 
$\approx$ 0.04). The result of a spline fit to 
\hbvbc\ after \heii\ subtraction is shown for both spectra in Fig. 
\ref{fig2} with parameters detailed in Table 2.
 
The main observational results of this investigation are: (a) the 
almost complete disappearance of the \hbbc\  (on 2000 May 8 the 
\hbbc\ flux is $\approx \frac{1}{10}$ ~ of the original flux) in a 
reasonably luminous quasar with bolometric luminosity 
$\sim$10$^{45}$ \ergss\ and (b) the much smaller variation in the 
\hbvbc\ component (a factor $\approx$ 2).
 
\section{Implications \label{impl}} 
 
\subsection{Implication 1: Existence of a VBLR Associated With \hbvbc} 
 
The 1990 observation of the  \hb\ region in PG 1416--129 showed 
inflections between narrow (\hbnc), broad  and very broad  line 
components. The 2000 spectrum shows the almost complete 
disappearance of the broad component \hbbc\ made more evident by 
the lack of change in \hbnc. The bulk of the residual broad line 
emission is associated with the very broad component  \hbvbc, also  
seen in 1990. The 
robustness of the \hbvbc\ component to a dramatic continuum 
change that quenched \hbbc\ strongly suggests that the VBC 
represents a distinct emitting component. Similar arguments have 
been made previously based on variability and line inflection 
evidence (Corbin 1997a,b). The weakness of \feii\ emission in PG 
1416--129 suggests that the \hbvbc\ cannot be attributed to a 
red-shelf  of enhanced \feii\ emission (Korista 1992). Do all RL 
and RQ (pop B: with FWHM(\hb ) $\simgt$ 4000 km s$^{-1}$ 
see Sulentic et al. 2000a) sources show a VBC? The answer is ``no'', at least 
for a VBC like the one observed in PG 1416--129, within current 
detection limits. Sources exist with no hint of a BC/VBC 
inflection, \hb\ red wing, or profile shift. 
 
Evidence also exists for VBC components of \heii\ (e.g., Ferland 
Korista \& Peterson 1990; Marziani \& Sulentic 1993) and \civ\ 
\AA\ (Laor et al. 1994; 1995) in several sources. It is very 
likely that these VBC components are related to the \hbvbc, 
although \heii\ may often show a more symmetric, boxy profile (as 
in PG 1138+222; Marziani \& Sulentic 1993). However, even if it 
is possible to disentangle the \heii\ emission from the 
surrounding \feii\ contamination, it is usually weak, with a flux 
distributed over a wide wavelength range often merging smoothly 
with the blue wing of \hb. \heii\ asymmetry measurements are 
therefore intrinsically difficult and  dependent on assumptions 
about the \hbbc\ shape even if the s/n of the data is high. Note 
that the classical BLR of \civ\ and other high ionization lines is 
called VBLR in the ILR/VBLR scheme followed by Brotherton et al.  
(1994) (see Sulentic \& Marziani 1999). In Corbin (1997ab) ILR= 
our NLR+BLR but VBLR definition is the same. The \civ\ VBC that we 
refer to is redshifted and much broader than the BC in analogy to 
\hb\ (see e.g. Laor et al. 1994, 1995). 
  
\subsection{Implication 2: Physics of the BLR and VBLR \label{phys}} 
 
\paragraph{BLR} The light travel time $\tau_{\rm LT}$ for a 
spherically symmetric BLR is $\tau_{\rm LT} \approx 500 \rm 
L^{\frac{1}{3}}_{H\beta,42} f_{\rm f,-7}^{-\frac{1}{3}} 
n_{11}^{\frac{2}{3}}$~ days, where $\rm f_{f,-7}$ ~  is the 
filling factors of the line emitting gas in units of 10$^{-7}$, 
$\rm L_{H\beta,42}$~ is the \hb\ luminosity in units of 10$^{42}$ 
ergs s$^{-1}$, and n is the number density in units of 10$^{11}$ 
\cm3. The \hbbc\ luminosity of PG 1416--129 is L$_{H\beta_{\rm 
BC}} \approx 8\times 10^{41}$ \ergss\ (excluding VBC emission; 
H$_0$ = 75 \kms Mpc$^{-1}$, q$_0$ = 0.5). The $\tau_{\rm LT}$\ for 
the BLR of PG 1416--129 is therefore most likely between several 
months and a few years. The almost complete disappearance of the 
\hbbc\ in $\simlt$10 yr does not pose any fundamental challenge. 
The simplest scenario is that \hbbc\ faded following a high 
amplitude (a factor $\simgt$ 4) decrease in ionizing continuum 
level.  The \hbbc\ luminosity can be  conventionally explained as 
due to an ensemble of optically thick clouds of very small 
filling factor ($\rm f_{\rm f} \sim 10^{-7}$) in a spherically 
symmetric region extending from the VBLR radius up to 10$^{19}$ 
cm. 
 
Another potentially very important constraint arises from the 
apparent absence of a BC in the \heii\ profile of PG 1411-129.  
Examination of the PG spectral 
atlas (BG92) indicates that when a broad component of \heii\ is 
visible, it  is usually broader than \hbbc. This is especially 
true for population B and RL sources (e.g. Corbin \& Smith 2000). 
In PG 1416--129 there is evidence for a \heiivbc\ but only in our 
new spectrum where \hbbc\ has almost disappeared. Negligible 
\heii\ emission (I(\heiibc)/I(\hbbc) $\simlt$ 0.03) implies that 
the ionization parameter must be very low, $\Gamma \simlt $ 
10$^{-4}$ in the BLR of PG1416--129. This condition is satisfied 
if the emitting gas is either high density (\ne $\simgt$ 10$^{12}$ 
\cm3) or located farther away  from the continuum source (a 
condition not supported by the line  width and by the variability 
timescale). If most \heii\ is  produced in a VBLR rather than a 
BLR, the problem of an ionizing photon deficit (Korista et 
al. 1997) may vanish  and we may need to reconsider the formation 
of the high ionization lines (HIL) in Pop B and RL AGN. 
 
\paragraph{VBLR} Several lines of evidence (Corbin 1997a,b; Sulentic 
et al 2000b) suggest that a very broad line region (VBLR) of 
optically thin gas exposed to a very strong radiation field is 
located at the inner edge of the BLR.  In the case of PG1416--129 
the empirical evidence includes: (1) a difference of 9000 km/s 
between FWHM(BC) and FWHM (VBC) and (2) a much weaker response by 
the \hbvbc\ to a large continuum change. In an optically thick 
medium the intensity of a recombination line  is governed by the 
luminosity of the ionizing continuum. If the medium is optically 
thin the intensity of the same recombination line is governed by 
the volume and density of the cloud distribution and is not 
directly related to the luminosity of the ionizing continuum. The 
much larger decline of \hbbc\ with respect to \hbvbc\ can 
therefore be explained if \hbbc\ is emitted in an optically thick 
medium, while a significant fraction ($\simgt$ 50\%) of \hbvbc\ is 
emitted by optically thin gas (see e.g. Shields et al. 1995). 
 
In the case of PG 1416--129, photoionization calculations 
performed using CLOUDY 94 (Ferland 2000) are able to reproduce 
the \hbvbc\ luminosity assuming that an optically thin screen  of 
gas surrounds the continuum sources at a distance of 10$^{17.5}$ 
cm, with $n_e \sim 10^{11}$ \cm3, and $\Gamma \simgt$ 10$^{-2}$. 
The \hbvbc\ is  strongly redshifted and redward asymmetric (in PG 
1416--129 the peak shift $\rm \Delta v_r \approx 850$\kms, while 
the shift at line base is $\rm \Delta v_r \approx 4500$ \kms). If 
we ascribe the shift to gravitational plus transverse  redshift 
(Corbin 1997b), then it is  $\rm \Delta v_r \approx 130 \rm 
M_{9,\odot} r_{18}^{-1} $ \kms, and  a black hole mass of a few 
10$^9$ \msol\ is needed to produce $\rm \Delta v_r \approx 1000$ 
\kms at r$\approx$10$^{17.5}$ cm. 

Any optically thin screen with $\Gamma \simgt 0.01$\ is 
expected to be a strong source of \heii; for $\Gamma \sim 0.1$ 
\heii\ can become comparable to \hb. In the case of PG 1416--129 
milder conditions seem to be appropriate. We measure an upper limit of
I(\heiivbc)/I(\hbvbc) $\approx$ 0.25 (Table \ref{tab1}), a value 
close to what has been held for a long time to be the canonical 
value for type 1 AGN. This is consistent with $\Gamma 
\approx 10^{-2}$ deduced for the optically thin shell covering 
all the source needed to explain the luminosity of \hbvbc.

\subsection{Implication 3: An RL -- RQ Population B Connection?} 
 
We have recently described an Eigenvector 1 parameter space as the 
optimal discriminator between various AGN broad line classes 
(Sulentic et al. 2000a,b). Sources with FWHM (\hbbc) $\simlt$ 4000 
\kms\ (Population A) tend to be RQ while broader lined sources 
are much more often RL. We identified a RQ Population B that shows 
line profile and soft X-ray properties indistinguishable from, 
especially flat spectrum, RL sources. Most of the  RL and 26\%\ of 
the RQ sources in the PG sample (BG92) fall in the population B 
domain. The various planes of the correlation space show: (1) 
reasonably strong correlations among pop A sources and (2) little or no 
correlation among the RL and RQ POP B sources, although their 
mean values are an extension of the pop A correlations (Sulentic et 
al. 2000a). 
 
Observational commonalities (and at the same time  differences from Pop 
A) between RL and RQ Pop B sources include: (i) stronger and more 
frequent optical variability (Ulrich et al. 1997), (ii) more 
complex/boxy/asymmetric profiles (Sulentic 1989; Eracleous \& Halpern 
1994; Marziani et al. 1996), (iii) occurrence of double-peaked profiles 
(e.g. Chen et al. 1989; Sulentic et al. 1995), (iv) occurrence of large 
single-peaked red/blue line shifts ($\simgt$10$^3$ \kms) and asymmetries
(e.g. Marziani et al. 1993; Gaskell 1983), (v) the absence of a 
systematic \civ\ blueshift (Marziani et al. 1996; Sulentic et al. 2000a), 
(vi) the absence of a soft X-ray excess (Yuan et al. 1998; Sulentic 
2000a) and (vii) the presence of a VBLR emitting component in an uncertain 
number of sources. All line related comments except (v) refer to 
the Balmer lines where the phenomenology is better established. 
 
The VBLR property appears to be an important commonality between 
RL and RQ pop B.  Sources like B2 1721-34/PKS1101-32 (Corbin 
1997ab), PKS0837-12 (Corbin \& Smith 2000), PKS0454-22 (Corbin \& 
Boroson 1996), PKS0214+10 (Eracleous \& Halpern 1994), PKS 
0403-132/ PKS0405-123 (Marziani et al. 1996), 0159-117 
(Brotherton 1996) are virtual twins of the 1990 spectrum of PG 
1416--129. Detection and accurate measurement of the VBLR feature 
requires at least moderate resolution ($\simlt$5$\rm{\AA}$) and 
s/n ($\simgt$20 in the continuum near \hb. RQ pop B analogs of 
PG1416--129 can be found in Marziani et al. (1996) where even 
relatively \feii\ strong RQ source Fairall 9 shows a VBLR 
feature. PG1416--129 shows us that pop B RQ sources can have a 
VBLR as strong as the strongest examples among RL AGN. 
 
The situation is less clear for much \feii\ stronger Population A 
sources.  \hbbc\ is in several cases observed to be blue-ward 
asymmetric, where the stronger blue wing may be associated with a 
high ionization wind emitting most of \civ\ (Marziani et al. 
1996; Sulentic et al. 2000ab). The available evidence suggests 
that RQ Pop A sources lack a VBLR like the one observed in Pop B 
objects (i.e. redshift$\sim$10$^3$ km/s and FWHM$\sim$10$^4$ 
km/s). Careful analysis of the \feiiq-contaminated \heii\ line 
profile would help to clarify the issue.

\subsection{Implication 4: Do We Always Measure the Same ``BLR'' in AGN?} 
 
PG1416--129 suggests that the answer to this question is ``no''. 
Sources with ``naked'' VBLR lines certainly exist. A spectrum of 
PG1416--129 with lower s/n or resolution would not 
reveal the small BLR residual as a feature. In 1990 BG92 reported 
FWHM(\hbbc)  $\approx$ 6100 \kms\  and \rfe\ $\approx$ 0.2 for 
PG1416. The classical FWHM(\hbbc) was overestimated and \rfe\ 
underestimated. This mis-estimation of BLR properties will occur 
whenever a significant VBC is present unless the local continuum 
is set above that component. Using the 2000 spectrum to model the 
VBLR component in the 1990 spectrum (see Fig. \ref{fig2} and Table 
\ref{tab2}) suggests that the correct values for the FWHM(\hbbc) 
and \rfe\ are 4000 km/s and 0.7 respectively. These parameter 
changes certainly exceed published error estimates for such 
measures. In the Eigenvector 1 correlation plane of FWHM(\hbbc) 
vs. \rfe, this change will move PG 1416--129 towards the 
population A  domain. It will also produce an apparently \feii\ 
``stronger" source (from \rfe $\approx$0.2 to 0.7) than is 
typically found for broad line RL/RQ pop B sources. 
 
If we parameterize PG1416--129 today we would measure FWHM 
(``H$\beta_{BC}$'') $\approx$ 13000 \kms\ along with  very 
weak/undetectable \feii. We expect no correlation between those 
two parameters because we know that \feii\ strength correlates 
with the classical \hbbc\ and {\em not} with the \hbvbc.  We would 
in fact be measuring a ``naked'' VBLR. 
The implication for  statistical studies and sample comparisons 
is clear: we must take into account the VBLR feature or we are 
not measuring the same thing in different sources. Sometimes we 
measure a pure BLR component, sometimes a pure VBLR component and 
sometimes a composite. The implications for the Eigenvector 1 
space are also clear. It is possible that the pop A-B differences 
motivated by the PG sample reflect the presence of the extra VBLR 
emitting component in Pop B sources, and that the classical BLR 
may be more similar from source to source than we have 
appreciated. 
 
\subsection{Implication 5: Inferences from Variability/Reverberation Studies} 
 
Much of the predictive power of variability/reverberation studies 
involves determining the sequential response of the line core and 
wings. These results have previously been argued to 
support dominance of radial or rotational motions in the BLR (e.g. Gaskell 
1988; O'Brien et al. 1998; Goad et al. 1999). 
The implication of the VBLR is that the red wing of the 
line may have nothing to do with the line core.  
Therefore the presence or lack of a pre-, same or post response  
allows us to infer nothing about the source structure. Monitoring  
data for sources like Fairall 9 could now be profitably  
reprocessed in the light of the VBLR concept.

In a sample of 14 RL sources involved in an annual spectroscopic 
campaign Corbin \& Smith (2000) report that most changes occur in 
the core of the line rather than the wings. They find that B2 
1721-34 shows a decrease in the core component as the continuum 
declined. They also argue that the wings of the line have 
increased at the same time. Close examination of their spectra 
show that most of the change in the wings has occurred on the 
blue side. All of this is consistent with the idea that the peak 
is dominated by the classical BLR which is optically thick. The 
red wing, especially, arises in large part in optically thin gas, 
so it would be expected to change much less. In any case the two 
wings may be phenomenologically distinct. 
 
It is important to stress that AGN Pop A sources mirror the behavior of 
Pop B and RL AGN. The blue wing of \hb\ seems to be associated with an 
optically thin wind emitting mostly HIL like \civ. The 
blue wing in population A sources may therefore show a lack of response  
to continuum change but for reasons unrelated to a VBLR component.

\section{Conclusion} 
 
The quasar PG1416-129 shows two broad line regions. The  
classical BLR and  VBLR are identified through inflections in the  
low ionization Balmer line profiles. They have a different place  
of origin and also arise from a physically distinct (opticaly  
thick/thin) regions as indicated by the difference of their response  
to the continuum change over the past ten years.

\acknowledgments

MC, PM and JS acknowledge support from the Italian Ministry of University and 
Scientific and Technological Research (MURST) through grant Cofin 98-02-32. 
JS and TZ acnowledges support and telescope time from IA/UNAM. TZ acknowledges 
support from the Slovene Ministry of Research and Technology.

\begin{deluxetable}{lccc} 
\tablecolumns{4} \tablewidth{0pc} 
\tablecaption{ PG 1416--129 
 Emission Line Spectrum\tablenotemark{a}\label{tab1}} \tablehead{ & 
\colhead{Flux\tablenotemark{b}} & \colhead{W\tablenotemark{c}} & 
\colhead{FWHM\tablenotemark{d}} \\ 
& & \colhead{[\AA]} & \colhead{[km s$^{-1}$]}} 
 
\startdata 
H$\gamma_{\rm NC}$ & 1.1\tablenotemark{e} & 10:\tablenotemark{e} & $\simlt$ 400\tablenotemark{e}  \\ 
 
H$\gamma_{\rm BC+VBC}$& 11.4:\tablenotemark{f} & 80:\tablenotemark{f} & $\simgt $ 4500  \\ 
 
[\ion{O}{3}]$\lambda$4363 & 2.2:\tablenotemark{e} & 20:\tablenotemark{e} 
& 1000:\tablenotemark{e} \\ 
 
\ion{He}{2} $\lambda$4686$_{\rm NC}$& 0.6:\tablenotemark{g} & 1.0:\tablenotemark{g} & 850\tablenotemark{g}\\ 
\ion{He}{2} $\lambda$4686$_{\rm VBC}$& 10.0\tablenotemark{h} & 75\tablenotemark{h}& 14000\tablenotemark{h} \\ 
 
H$\beta_{\rm NC}$ & 2.0 & 14 & 250\\ 
 
H$\beta_{\rm BC+VBC}$ & 40.0 & 300 & 9000 \\ 
 
[\ion{O}{3}]$\lambda$4959 & 9.0  & 68 & 370 \\ 
 
[\ion{O}{3}]$\lambda$5007 & 27.7& 205  & 370\\

\enddata 
\tablewidth{\textwidth}
\tablenotetext{a}{Line parameters measured on the 2000 May 8 
spectrum unless otherwise noted, after rescaling to the 
[\ion{O}{3}]$\lambda$4959,5007 fluxes of the 1990 spectrum.} 
\tablenotetext{b}{Observed, in units of 10$^{-15}$ ergs s$^{-1}$ 
cm$^{-2}$.} 
\tablenotetext{c}{Rest frame equivalent widths are 
measured  against the bare continuum.} 
\tablenotetext{d}{Corrected for instrumental profiles assuming 
FHWM$_{instr} \approx 7$ \AA\ and 3.5 \AA\ for the 1990 and 2000 
spectrum respectively.} \tablenotetext{e}{Measured after 
subtraction of a scaled and shifted H$\beta_{\rm BC+VBC}$\ to 
mimick H$\gamma_{\rm BC+VBC}$.  }\tablenotetext{f}{Measured 
scaling and shifting H$\beta_{\rm BC+VBC}$.} 
\tablenotetext{g}{Measured on the 1990 Feb. 17 spectrum.} 
\tablenotetext{h}{Measured using \hbvbc\ shifted and scaled to 
1/4 its original flux.} 
\tablenotetext{\ }{":" : uncertainty higher than standard estimate (see text).}
\end{deluxetable} 
\begin{deluxetable}{lccccccc} 
\tablecolumns{4} \tablewidth{0pc} \tablecaption{PG 1416--129 
H$\beta$ Emission Line Variability\label{tab2}} \tablehead{ 
\colhead{Line Identification} & \multicolumn{3}{c}{1990 Feb. 
17\tablenotemark{a}} && \multicolumn{3}{c}{2000 May 
8\tablenotemark{b}} \\ 
\cline{2-4} \cline{6-8} \\ 
& 
 \colhead{Flux}   & \colhead{W}  & \colhead{FWHM}& & \colhead{Flux}   & 
 \colhead{W}  & \colhead{FWHM} \\ 
 & & [\AA] & [km s$^{-1}$] &&& [\AA] & [km s$^{-1}$] } 
\startdata 
H$\beta_{\rm BC+VBC}$&  86.0   & 160 &  6000  && 40.0 & 300 & 9000   \\ 
H$\beta_{\rm BC}$    &  23.0   &  47 &  4000    && 2.0  & 13 & 1450   \\ 
H$\beta_{\rm VBC}$   &  63.0   & 110  & 13000   && 38.0 & 220 & 13000  \\ 
\enddata 
\tablenotetext{a}{Specific flux at 4500 \AA\ F$_\lambda \approx$ 
7.5$\times10^{-16}$ ergs s$^{-1}$ cm$^{-2}$ \AA$^{-1}$.} 
\tablenotetext{b}{Specific flux at 4500 \AA\ F$_\lambda \approx$ 
1.9$\times10^{-16}$ ergs s$^{-1}$ cm$^{-2}$ \AA$^{-1}$.} 
 
\end{deluxetable}

\end{document}